**Deciphering *SCN2A*: A comprehensive review of rodent models of *Scn2a* dysfunction**

Katelin E.J Scott [1,3,4,†], Maria F. Hermosillo Arrieta [2,3,4,†], Aislinn J. Williams [3,4,†,*]

[1] Interdisciplinary Neuroscience Graduate Program
[2] iDREAM
[3] Iowa Neuroscience Institute
[4] Department of Psychiatry
[*] Correspondence: Dr. Aislinn Williams, aislinn-williams@uiowa.edu
[†] University of Iowa, Iowa City, Iowa

**Background**

*SCN2A* related disorders are highly heterogeneous and manifest in a variety of diagnoses including self-limited familial and non-familial infantile epilepsy (SeLFNIE) (previously benign familial infantile seizures or BFNIS) (Lauxmann et al., 2018), epileptic encephalopathies (EE) (Adney et al., 2020; Berecki et al., 2018; Kamiya et al., 2004; J. Li et al., 2016; Liao et al., 2010; Ogiwara et al., 2009; Shi et al., 2012), infantile spasms (Sundaram et al., 2013), ataxia (Murray et al., 2023), autism spectrum disorder (ASD) (Ben-Shalom et al., 2017; Sanders et al., 2018; Weiss et al., 2003), intellectual disability (ID) (Begemann et al., 2019; J. Li et al., 2016), and schizophrenia (Carroll et al., 2016; Fromer et al., 2014; J. Li et al., 2016; Suddaby et al., 2019). These disorders have been associated with functional alterations. *SCN2A* encodes the alpha subunit of the voltage-gated sodium channel Na$_v$1.2, which is involved in action potential initiation and backpropagation in glutamatergic neurons (Hu et al., 2009; Spratt et al., 2019).

Researchers have attempted to elucidate the complexity of *SCN2A*-related disorders through non-human mammalian models. This review aims to evaluate and compare the published rodent models to consolidate findings, identify limitations, and highlight future research directions (Supplemental Table 1). Thus far, over 30 *Scn2a* mouse models have been engineered and published, and there are at least 2 other rodent models, beyond mice, available for *SCN2A* research. For a review of non-mammalian models and mechanisms of *SCN2A*-Related Disorders, please see Hedrich et al., 2020 and Kruth et al., 2020.

## Mouse Models of *Scn2a* Insufficiency

### First *Scn2a* Constitutive Haploinsufficient Mouse Model

The first published *Scn2a* mouse model was created by disrupting the first 89 amino acids in exon 1 of the *Scn2a* gene via homologous recombination (Planells-Cases et al., 2000). This mouse model revealed that homozygous knockout of *Scn2a* is perinatally lethal in mice. The *Scn2a* haploinsufficient (*Scn2a$^{+/-}$*) mouse model was viable and able to reproduce. Since 2000, the Planells-Cases model has been widely used and robustly characterized (Indumathy et al., 2021; Léna & Mantegazza, 2019; Marcantonio et al., 2023; Middleton et al., 2018; Misra et al., 2008; Miyamoto et al., 2019; Nelson et al., 2024; Ogiwara et al., 2018; Spratt et al., 2019, 2021; Tamura et al., 2022; Tatsukawa et al., 2019; C. Wang et al., 2024). It is likely the most extensively characterized model of *SCN2A*-related disorders.

Although robustly characterized, there are some conflicting behavioral results from the various studies utilizing the model. For instance, Tatsukawa et al., 2019 found that *Scn2a$^{+/-}$* mice were hypersocial compared to their wildtype littermate controls and Marcantonio et al., 2023 identified a similar phenotype in juvenile female *Scn2a$^{+/-}$* mice. However, Léna and Mantegazza, 2019, found mild social behavior deficits in juvenile male *Scn2a$^{+/-}$* mice that attenuated with age. Some studies also found no differences between the sociability of *Scn2a$^{+/-}$* mice and their littermate controls (Indumathy et al., 2021; Spratt et al., 2019). A summary of general behavioral findings from the (Planells-Cases et al., 2000) *Scn2a$^{+/-}$* mouse model is provided in Figure 1 and Supplemental Table 2.

When first characterized, it was thought that *Scn2a$^{+/-}$* mice did not experience seizures (Misra et al., 2008; Planells-Cases et al., 2000), but recently, electrographic spontaneous absence-like seizures have been reported in *Scn2a$^{+/-}$* mice (Miyamoto et al., 2019; Ogiwara et al., 2018). The authors suggested that the "outwardly healthy appearance" of the mice could be why past seizure activity had been undetected (Miyamoto et al., 2019). Now there are also reported cases of this model having convulsive seizures: Schamiloglu et al.,2023 identified 16% of *Scn2a$^{+/-}$* mice actively seizing at the end of a dynamic foraging behavior task. The mice used in the task had restricted water consumption before testing, which could have induced the seizures (Schamiloglu et al., 2023). This raises the possibility that the severity and type of seizure endured by *Scn2a$^{+/-}$* mice can be modulated by experience.

Beyond seizures, the Planells-Cases *Scn2a$^{+/-}$* mice have been found to have slower and more broad action potential generation, at least in the first postnatal week (Miyamoto et al., 2019; Ogiwara et al., 2018; Spratt et al., 2019; Tamura et al., 2022). Additionally, immature glutamatergic cortical neurons from *Scn2a$^{+/-}$* mice display decreased neuronal excitability, but mature neurons from these animals are hyperexcitable (Spratt et al., 2019, 2021). These findings indicate that impairments in action potential depolarization and cell excitability may depend on developmental stage. Additionally, partial loss of Na$_v$1.2 impaired cortical-striatal firing and may contribute to the identified seizures through maladaptive changes in the cortical-striatal circuit (Hu et al., 2009; Ogiwara et al., 2018). Haploinsufficiency of Na$_v$1.2 also impaired dendritic excitability and backpropagation, which can further alter neuron repolarization and activity (Nelson et al., 2024). Overall, constitutive *Scn2a* haploinsufficiency appears to impair excitatory cortical neurons in a developmental stage-specific manner and increases seizure susceptibility, although the precise mechanism linking these changes remains unknown.

**Other Constitutive Haploinsufficient Mouse Models**

Additional constitutive haploinsufficient *Scn2a* mouse models exist. While not as well characterized as the Planells-Cases model, they provide further insight into the behavioral and physiologic effects of various genetic disruptions inducing Na$_v$1.2 haploinsufficiency.

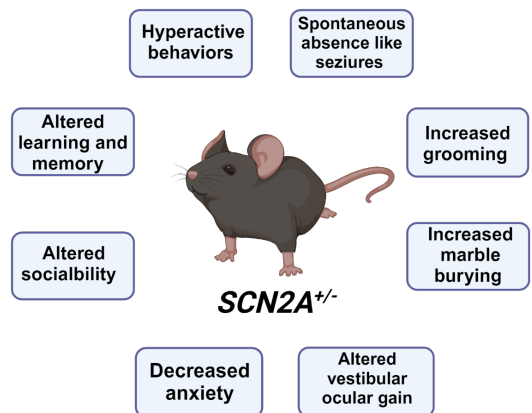

**Figure 1:** Summerized main behavioral findings from the *SCN2A* $^{+/-}$ mouse model. See a more detailed summary of behavioral experiments in Supplemental table 2. Created in BioRender. Williams lab, A. (2024) https://BioRender.com/b81m903

**Na$_v$1.2$^{adult}$**

**~ 2months and older (male only)**
- Social
  - 16x more likely to jump out of social interaction box
- Anxiety and Hyperactivity
  - Elevated Plus Maze
    - Reduced anxiety
- Seizures
  - More susceptible to PTZ-induced seizures

**Scn2a $^{fl/+}$**

**2 to 4 months (Male only)**
- Social
  - Direct Social Interaction Test
    - Increased direct social interation
- Learning and Memory
  - Fear conditioning
    - Increased Fear memory
  - Spatial Object Recognition
    - Reduced Spatial learning and memory
- Motor
  - Decreased locomotion in familar environments
- Anxiety and Hyperactivity
  - Open Field Test
    - Mild light induced hypoactivity

**4 days to 1 month (Male only)**
- Social
  - Direct Social Interaction Test
    - Slight increase direct social interaction
- Motor
  - Open Field Test
    - Decreased locomotion

**Scn2a$^{+/-}$; Kcna1$^{+/-}$**

**~ 1 month**
- Innate and Repetitive Behaviors
  - Reduced marble burying

**Scn2a$^{+/-}$; Kcna1$^{-/-}$**

**~ 1 month**
- Social
  - 3-chamber social interaction heading typo
    - Reduced social novelty sniffing
- Repetitive and Innate Behaviors
  - Reduced nestlet shredding and marble burying

**2 to 3 months**
- Seizures
  - Scn2a$^{+/-}$ decreases spontaneous seizure duration in Kcna1$^{-/-}$ mice

**Figure 2:** Summary of main behavioral findings from other constitutive haploinsufficient *Scn2a* mouse models. Model constructs not included in this summary table either did not evaluate behavior or did not identify results in their *Scn2a* model construct distinct from wildtype animals. See a more detailed summary of behavioral assessment in these animals in Supplementary table 2. (PTZ= pentylenetetrazole)
Created in BioRender. Williams lab, A. (2024) https://BioRender.com/t38r001

In one such model, the adult Na$_v$1.2 channel isoform was expressed in neonatal animals using homologous recombination (Gazina et al., 2015). A targeting construct leading to the desired mutation was injected into blastocysts of C57BL/6 mice to obtain chimeric mice, which were then bred with wildtype C57BL/6 mice (Gazina et al., 2015). The resulting offspring were then crossed with a cre-transgenic mouse strain B6.C-Tg(CMV-Cre)1Cgn/J to remove the neomycin cassette and thus generate a stable and constitutive line of heterozygous Na$_v$1.2$^{adult/+}$ mutants, simply referred to as Na$_v$1.2$^{adult}$ mice (Gazina et al., 2015). The neonatal and adult isoforms of *Scn2a* are only one nucleotide different from one another in exon 5; the neonatal isoform (5N) expresses an asparagine at residue 209, while the adult isoform (5A) expresses aspartic acid at the same position (Gazina et al., 2010, 2015; Kruth et al., 2020). 5N is less excitable than 5A, and when 5A was expressed in neonatal mice, they demonstrated signs of hyperactivity and seizure activity (Figure 2, Supplemental Table 2) (Gazina et al., 2015). Data from the human channel also supports the idea of different electrophysical profiles between the two Na$_v$1.2 isoforms. Some gain of function human variants cause the Na$_v$1.2 5N isoform to mimic the electrophysical properties of 5A, leading to seizures (Muller, 2020; Thompson et al., 2020). The change in electrophysical properties of Na$_v$1.2 during the neonatal period may explain why these individuals had seizure profiles in infancy that attenuated with age (Muller, 2020; Thompson et al., 2020). Since the neonatal isoform of *Scn2a* has different electrophysical properties than the adult isoform and can lead to different phenotypic effects when altered, Na$_v$1.2 5N and 5A need to be further investigated to elucidate the functional roles of both in *SCN2A*-related disorders.

Another constitutive model deleted exons 4-6 to induce Na$_v$1.2 haploinsufficiency (*Scn2a $^{fl/+}$*) (Shin et al., 2019). Although a large part of the coding sequence for *Scn2a* was removed, including exon 5 that encodes for the 5N and 5A isoforms, about 60% expression of *Scn2a* was maintained in *Scn2a$^{fl/+}$* mice (Shin et al., 2019). *Scn2a$^{fl/+}$* mice did not display spontaneous seizure activity but did display decreases in neuronal activity and excitatory synaptic transmission (Shin et al., 2019). Additionally, a close electrophysical investigation of hippocampi from these mice revealed that the Schaffer collateral-CA1 pathway of these animals was typical; however, long-term potentiation was suppressed without affecting long-term depression through mechanisms independent of NMDAR-mediated synaptic transmission (Shin et al., 2019). Furthermore, *Scn2a*$^{fl/+}$ had impairments in reversal learning but displayed no measurable changes in grooming and anxiety-like behaviors (Shin et al., 2019)(Figure 2, Supplemental Table 2). It is important to mention that although no changes in anxiety-like behavior were detected, juvenile *Scn2a*$^{fl/+}$ mice displayed less overall locomotion in the open field test than their wildtype counterparts. This finding could further highlight a potential difference in the role of Na$_v$1.2 in juvenile vs adult animals (Shin et al., 2019). Furthermore, RNA scope data from the *Scn2a*$^{fl/+}$ mice indicate that *Scn2a* mRNA is expressed in both GABAergic and glutamatergic neurons of the neocortex and hippocampus in adult animals, suggesting that *Scn2a* could also function and exist in inhibitory cells.

A similar construct, the *Scn2a*$^{+/KI}$ model, was generated by deleting exons 3-5 of one *Scn2a* allele and inserting an in-frame eGFP sequence to visualize cells that expressed *Scn2a* (Tamura et al., 2022). Seizures were induced in these animals with pentylenetetrazole, but no difference was identified in susceptibility to seizures between *Scn2a*$^{+/KI}$ and wildtype C57BL/6J mice (Tamura et al., 2022). These additional constitutive *Scn2a*$^{+/-}$ models indicate that although all express between 50% to 60% of native *Scn2a* protein, the effect of each manipulation does not have an identical impact on phenotypic presentation.

There is also at least one model that paired the Planells-Cases *Scn2a*$^{+/-}$ mouse with a *Kcna1*$^{+/-}$ mouse to generate mice heterozygous for both genes or heterozygous for *Scn2a* but null for *Kcna1* (Mishra et al., 2017). *Kcna1* encodes for K$_V$1.1, a voltage-gated potassium channel, and is a monogenetic risk gene for epilepsy. The mixed model enabled a close investigation of how *Scn2a* and potassium channels interact, which is discussed in more detail later in this manuscript. Both *Scn2a*$^{+/-}$;*Kcna1*$^{+/-}$ mice and *Scn2a*$^{+/-}$;*Kcna1*$^{-/-}$ animals had attenuated repetitive behaviors compared *Scn2a*$^{+/-}$ mice and were more similar to wildtype C57BL/6J mice (Indumathy et al., 2021). It has also been noted that the seizure phenotype apparent in *Kcna1*$^{-/-}$ was attenuated in *Scn2a*$^{+/-}$;*Kcna1*$^{-/-}$ mice (Mishra et al., 2017). The improvements in behavior and seizure phenotype of *Scn2a*$^{+/-}$;*Kcna1*$^{+/-}$ and *Scn2a*$^{+/-}$;*Kcna1*$^{-/-}$ suggest that Na$_v$1.2 and Kv1.1 may be therapeutic targets for one another and are likely playing a compensatory role for each other when one is impaired. When both channels are partially lost, there appears to be a synergistic positive effect on overall health outcomes, likely because the channels cannot overcompensate for one another and cause maladaptive changes.

Lastly, another haploinsufficient *Scn2a* mouse model was generated by creating a point mutation at amino acid 38, which switched Lys for Gln (*Scn2a*$^{K38Q}$) (Kotler et al., 2023). This point mutation removed the only SUMO-conjugation site in Na$_v$1.2 channels (Kotler et al., 2023). This model helped reveal that SUMOylation of Na$_v$1.2 channels prolongs the decay time constant of EPSPs greater than 10 mV in amplitude, but this effect is lost in *Scn2a*$^{K38Q}$ mice. Furthermore, SUMOylation was found to impact the speed of forward and backpropagating action potentials of cortical pyramidal cells from *Scn2a*$^{K38Q}$ mice (Kotler et al., 2023). These findings indicate that changes in SUMOylation and possibly other post-transcriptional modifications influence the observed changes in action potential back-propagation observed in *Scn2a* mutant mice.

**Hypomorphic *Scn2a* mouse models**

Most *Scn2a* constitutive loss of function (LOF) mouse models result in *Scn2a* haploinsufficiency and retain around 50% of Na$_v$1.2 channel expression. However, there are at least two constitutive *Scn2a* models that result in less than 50% of Na$_v$1.2 channel expression; a severe hypomorph and a gene trap model(C. Wang et al., 2024). These models both retain around 25% or less of Na$_v$1.2 channel expression. As global homozygous loss of Scn2a results in perinatal death, these models have been instrumental in exploring the consequences of severe loss of *Scn2a* without complete deletion.

The severe *Scn2a* hypomorphic mouse was engineered using Crispr-Cas9 to induce a 27-base pair frameshift in the C-terminal region of the *Scn2a* protein creating *Scn2a*$^{\Delta1898/+}$(H.-G. Wang et al., 2021), which resulted in ≤25% or less of native Na$_v$1.2 (H.-G. Wang et al., 2021). *Scn2a*$^{\Delta1898/+}$ mice displayed hyperactivity and reduced anxiety-like behaviors, which is similar to other *SCN2A* LOF models (Figure 3, Supplemental Table 2) (Gazina et al., 2015; Léna & Mantegazza, 2019; Tatsukawa et al., 2019; H.-G. Wang et al., 2021). Additionally, *Scn2a*$^{\Delta1898/+}$ mice displayed increased social behaviors, which aligns with observations from individuals with *SCN2A*-related disorders (Figure 3, Supplemental Table 2)(Sanders et al., 2018). Cultured excitatory cortical pyramidal neurons from *Scn2a*$^{\Delta1898/+}$ mice displayed both reduced voltage-gated Na+ channel currents and reduced neuronal excitability, but inhibitory neurons were unchanged (H.-G. Wang et al., 2021). Brain slices from these animals further revealed reduced excitatory synaptic input onto cortical pyramidal neurons (H.-G. Wang et al., 2021).

A *SCN2A* model created using a Tm1a trapping cassette (*Scn2a*$^{gtKO/gtKO}$) reduced Na$_v$1.2 expression to around 25% that of wildtype similar to the *Scn2a*$^{\Delta1898/+}$ mouse (Eaton et al., 2021; H.-G. Wang et al., 2021). The *Scn2a*$^{gtKO/gtKO}$ displayed significant alterations in innate behaviors like nest building and grooming (Figure 3, Supplemental table 2) (Eaton et al., 2021). These mice also had increased anxiety-like behaviors, which contrasts the behavioral findings seen in other models of *Scn2a* loss (Figure 3, Supplemental table 2) (Gazina et al., 2015; Léna & Mantegazza, 2019; Tatsukawa et al., 2019; H.-G. Wang et al., 2021). Adult *Scn2a*$^{gtKO/gtKO}$ mice were found to have increased neuronal excitability, which co-occurred with a higher voltage threshold in medium spiny neurons, which parallels the paradoxical hyperexcitability of cortical pyramidal neurons lacking Na$_v$1.2 (MSNs) (Spratt et al., 2021; Zhang et al., 2021). There was reduced expression and current outputs of some potassium channels in these mice, likely to compensate for the increased excitability induced by Na$_v$1.2 depletion ( Zhang et al., 2021). These results indicate that changes in neuronal excitability due to Na$_v$1.2 reduction are dependent on both levels of *SCN2A* expression and cell type. Regardless, the hyperexcitability of some neuron types in response to Na$_v$1.2 loss/reduction is likely contributing to the seizure disorders experienced by individuals with LOF *SCN2A* mutations (Zhang et al., 2021). Microglia in the *Scn2a* $^{gtKO/gtKO}$ mice have morphology changes distinct from wildtype C57BL/6N mice and may contribute to some of the noted phenotypes in these animals (Wu et al., 2024). The *Scn2a* $^{gtKO/gtKO}$ mice further had alterations in the core clock genes and the suprachiasmatic nucleus, leading to altered sleep patterns (Figure 3, Supplemental table 2) (Ma et al., 2022). This is the first model of *SCN2A* loss to evaluate sleep disturbances, although they are commonly reported in individuals with *SCN2A*-related disorders (Sanders et al., 2018). Alterations in sleep architecture and regulation should be further investigated in other *SCN2A* models to gain further insight into Na$_v$1.2's role in sleep. Improving the understanding of Na$_v$1.2's role in sleep can enhance the treatment of *SCN2A*-related sleep disorders by identifying affected sleep cycles and mechanisms, leading to more precise treatment targets.

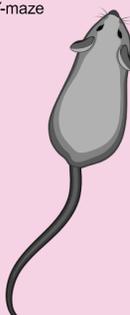

**Figure 3:** Summary of main behavioral findings from *Scn2a* hypomorphic models. For a more detailed summary please see Supplementary table 2. Created in BioRender. Williams lab, A. (2024) https://BioRender.com/x74k536

## Conditional and tissue specific deletion of *Scn2a* in mice

There are multiple *Scn2a* models that either reduce *Scn2a* levels at a specific age or induce tissue-specific changes in *Scn2a* expression. Many of these models utilize Cre-driver lines to induce conditional deletion of *Scn2a* in specific tissue and cell types (Gould & Kim, 2021; Miyamoto et al., 2019; Ogiwara et al., 2018; Spratt et al., 2019, 2021; Suzuki et al., 2023; Tamura et al., 2022; Tatsukawa et al., 2019; H.-G. Wang et al., 2021). Other models utilize unique approaches like short-hairpin RNAs and antisense oligonucleotides (M. Li et al., 2023; Nunes & Kuner, 2018).

Expressing shRNAs on postnatal day 21 enabled the deletion of Na$_v$1.2 in olfactory bulb granule cells (ObGC), which enabled the investigation of Na$_v$1.2's involvement in odor discrimination (Nunes & Kuner, 2018). Na$_v$1.2 channels are expressed along the cell surface and dendritic spines obGCs, and removal of *Scn2a* from obGCs was found to impair rapid and accurate odor discrimination (Figure 4, Supplemental Table 2) (Nunes & Kuner, 2018). Impaired odor discrimination was likely due to inhibition of synaptically connected mitral cells, diminished GABA release, and reduced granule cell spiking, which were all consequences of Na$_v$1.2 deletion in obGC (Nunes & Kuner, 2018). Currently, this is the only model to identify impaired odor discrimination in *Scn2a* mutant mice; investigations utilizing the Planells-Cases *Scn2a*$^{+/-}$ mouse have not revealed a difference in odor discrimination between mutant mice and their wildtype counterparts (Eaton et al., 2021; Léna & Mantegazza, 2019). These conflicting findings suggest that a threshold of $Na_v1.2$ is necessary for rapid and accurate odor discrimination.

Another unique method to reduce Scn2a expression relied on antisense oligonucleotides (ASO). An ASO was injected into the ventricles of 2 to 3-month-old Rbp4-Cre mice to both truncate *Scn2a* and create *Scn2a* insufficiency and selectively label L5 pyramidal neurons (M. Li et al., 2023). The *Scn2a* ASO mice displayed reduced sociability and increased hyperactivity and anxiety-like behaviors (Figure 4, Supplemental table 2). These mice also lost innate nest building, similar to data from *Scn2a $^{gtKO/gtKO}$* mice (Figure 4, Supplemental table 2) (M. Li et al., 2023). $Ca^{2+}$ transient currents were also reduced in *Scn2a* ASO mice, which decreased spontaneous cortical somatosensory neuronal firing and pairwise co-activation (M. Li et al., 2023). This study wanted to utilize the *Scn2a* ASO mouse model as a proxy for identifying the consequences of premature truncating variants (PTV). However, *Scn2a* haploinsufficiency is not generated until mice are between 2 and 3 months of age and neural networks have been fully established. Therefore, this model may not fully capture how prematurely truncating *SCN2A* can change neural circuits and signaling as the animal developed typically until induction and critical periods of $Na_v1.2$ involvement could have been missed.

Cre-driver lines have been commonly utilized to generate developmental stage, cell, and tissue-specific deletion or reduction of *Scn2a*. For instance, two Cre driver lines were utilized to delete *Scn2a* in either forebrain excitatory neurons *(Emx1-Cre)* or global deletion in inhibitory neurons (*Vgat*-Cre) in an *Scn2a* floxed animal (Ogiwara et al., 2018). Based on western blot data from P0.5 animals, *Scn2a$^{fl/fl}$Emx1-Cre* had a 30% reduction of $Na_v1.2$, while *Scn2a$^{fl/fl}$Vgat-Cre* resulted in around a 60% reduction of $Na_v1.2$ expression, further confirming that at least early postnatal mice express $Na_v1.2$ in both excitatory and inhibitory neurons (Miyamoto et al., 2019). Both *Scn2a$^{fl/fl}$Emx1*-Cre and *Scn2a$^{fl/fl}$Vgat-Cre* mice died within a few days of birth, likely due to homozygous loss of $Na_v1.2$ (Ogiwara et al., 2018). Western blot of *Scn2a$^{fl/+}$Emx1-Cre* and *Scn2a$^{fl/+}$Vgat-Cre* adult animals, however, revealed that *Scn2a$^{fl/+}$Emx1-Cre* had around 50% reduced $Na_v1.2$ expression in neocortex and hippocampus, but no significant reduction of $Na_v1.2$ was noted in *Scn2a$^{fl/+}$Vgat-Cre* mice, suggesting that in adult animals $Na_v1.2$ is mainly expressed in excitatory neurons (Ogiwara et al., 2018). Furthermore, *Scn2a$^{fl/+}$Emx1-Cre* and *Scn2a$^{fl/+}$Vgat-Cre* mice were viable, fertile, and outwardly healthy, although about a third of *Scn2a$^{fl/+}$Vgat-Cre* mice died unexpectedly around P18 - P25 of unknown causes (Ogiwara et al., 2018). The unexplained and sporadic early death in *Scn2a$^{fl/+}$Vgat*-Cre mice could point to $Na_v1.2$ having an important role in inhibitory cells early in development that is currently not understood. *Scn2a$^{+/-}$* mice have also been bred with a Vgat-venus line to fluorescently label inhibitory neurons to discern them from excitatory (Ogiwara et al., 2018). The Vgat-venus/*Scn2a$^{+/-}$* mice helped determine that inhibitory cells do not experience the same alterations in excitability demonstrated in glutamatergic excitatory neurons, which indicates that loss of $Na_v1.2$ impacts inhibitory and excitatory cells in distinct ways.

Behaviorally, *Scn2a$^{fl/+}$/Emx1-Cre* had significantly more rearing events in the open field assay than wildtype C57BL/6J littermates (Tatsukawa et al., 2019). *Scn2a$^{fl/+}$/Emx1-Cre* mice further displayed signs of possible hyperactivity and decreased anxiety-like phenotypes in the open field test. *Scn2a$^{fl/+}$/Vgat-Cre* mice did not display these phenotypes in the open field test, but in the elevated plus maze, *Scn2a$^{fl/+}$/Vgat-Cre* mice demonstrated reduced anxiety-like phenotypes (Figure 4, Supplemental table 2) (Tatsukawa et al., 2019). These results suggest that $Na_v1.2$ loss in dorsal-telencephalic excitatory neurons contributes to hyperactivity and repetitive behaviors, while anxiety-like phenotypes change in distinct ways when $Na_v1.2$ is lost in dorsal-telencephalic excitatory neurons or inhibitory neurons. $Na_v1.2$ expression in inhibitory neurons may be related to risk behavior, and loss of $Na_v1.2$ in these cells is what is driving the mice to explore more in the elevated plus maze. In contrast, $Na_v1.2$ in dorsal-telencephalic excitatory neurons may be more related to general exploration and locomotion, which is why there are different performances on the open field and elevated plus maze between the two conditional models.

ECoG data from *Scn2a*[fl/+]/*Emx1-Cre* mice indicates that these animals have absence-like seizure activity and these animals further displayed spike-wave discharges (SWDs) during moments of behavioral quiescence which is typical of seizure activity (Miyamoto et al., 2019; Ogiwara et al., 2018) (Figure 4, Supplemental Table 2). However, *Scn2a*[fl/+]/*Vgat-Cre* mice did not display seizure activity or behavioral quiescence (Miyamoto et al., 2019; Ogiwara et al., 2018). These data indicate that partial loss of Na$_v$1.2 in inhibitory neurons is unlikely to be driving seizures in these animals, but partial loss of Na$_v$1.2 in forebrain excitatory neurons is sufficient to cause seizures. Two other Cre-driver *Scn2a* models were engineered using the same *Scn2a* floxed mouse to elucidate if the SWDs had specific cortical striatal involvement (Miyamoto et al., 2019). The new lines, *Scn2a*[fl/fl]/*Trpc-Cre* mice and *Scn2a*[fl/fl]/*Ntsr-Cre* mice, deleted *Scn2a* from cortical layer 5 (L5) and cortical layer 6 pyramidal neurons (L6) respectively (Miyamoto et al., 2019). *Scn2a*[fl/fl]/*Trpc-Cre* mice, but not *Scn2a*[fl/fl]/*Ntsr-Cre* mice, displayed SWDs that matched the previously identified seizure phenotype, which indicates that L5 neurons and not L6 are likely involved in the identified absence seizure phenotype (Miyamoto et al., 2019). Further research will be needed to illuminate how each subclass of L5 neurons, intratelencephalic or extratelencephalic, is involved.

Bilateral injections of adeno-associated virus (AAV) expressing Cre recombinase were injected into *Scn2a*[fl/fl] mice to selectively delete *Scn2a* from mPFC or VTA to elucidate the role of Na$_v$1.2 in schizophrenia (Suzuki et al., 2023). When *Scn2a* was depleted in the mPFC, prepulse inhibition (PPI) was reduced; however, an increase in PPI response was found when *Scn2a* was depleted in the VTA (Figure 4, Supplemental Table 2) (Suzuki et al., 2023). Additionally, deletion of *Scn2a* in the mPFC was found to increase social interaction and anxiety-like behaviors but decrease locomotion when compared to control-treated animals; this effect was not seen when *Scn2a* was deleted in the VTA (Figure 4, Supplemental Table 2) (Suzuki et al., 2023). AAV-EF1α-Cre-mCherry has also been utilized to delete *Scn2a* in mPFC and label excitatory cells in *Scn2a*[fl/fl] mice to explore Na$_v$1.2 function without a specific focus on schizophrenia (Spratt et al., 2021). When Na$_v$1.2 is deleted from mature cortical pyramidal cells in the mPFC, the cells become paradoxically hyperexcitable (Spratt et al., 2021). However, these cortical neurons also had a hypoexcitable response for dendritic backpropagation, which impaired axonal repolarization (Spratt et al., 2021). Both the paradoxical hyperexcitability of the axon and the hypoexcitable dendritic backpropagation can be explained by Na$_v$1.2's unique role in regulating somatodendritic excitability through potassium channel priming (Spratt et al., 2019, 2021). Reduced dendritic backpropagation results in fewer potassium channels opening to repolarize the neuron and establish a refractory period, thus allowing for more action potentials while impairing repolarization. This phenomenon noted in Na$_v$1.2 deficient cells is likely one of the mechanisms in how LOF of Na$_v$1.2 can lead to seizure disorders but does not necessarily explain how the loss of Na$_v$1.2 contributes to other neuropsychiatric or developmental disorders. These studies highlight that although *Scn2a* likely has cell type, developmental stage, and brain region specific roles, the mechanism of how Na$_v$1.2 contributes to such a heterogeneous group of neuropsychiatric disorders and symptoms requires further investigation.

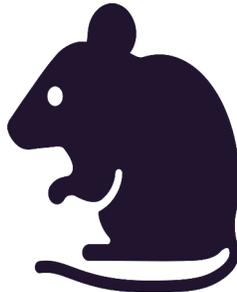

**Figure 4:** Summary of main behavioral findings from conditional and tissue specific deletion of *Scn2a* in mice models. Model constructs not included in this summary table either did not evaluate behavior or did not identify results in their *SCN2A* model construct distinct from wildtype animals. See a more detailed summary of behavioral assesment in these animals in Supplementray table 2. (VOR= Vestibular ocular reflex, SWD = Spike wake discharge) Created in BioRender. Williams lab, A. (2024) https://BioRender.com/c64k310

Another *Scn2a$^{+/fl}$* mouse under the CaMKIIα-Cre driver removed exon 2 of the *Scn2a* gene (Spratt et al., 2019) around P10. The *Scn2a$^{+/fl}$/CaMKII-Cre* mice developed with standard action potential threshold and spike output (Spratt et al., 2019). At P10, the animals start expressing cre in neocortical pyramidal cells and can then be manipulated for *Scn2a* knockdown (Spratt et al., 2019). If knockdown was induced around P10 by P50 peak dV/dt overlapped with constitutive *Scn2a$^{+/-}$* cells (Spratt et al., 2019). Ex-vivo culture of pyramidal cells from the model identified impaired dendritic excitability and action potential backpropagation even when haploinsufficiency was induced around P50 (Spratt et al., 2021). Additionally, conditional deletion of *Scn2a* in either parvalbumin or somatostatin-positive interneurons did not appear to impair excitability in these neurons, which is consistent with previous findings (Ogiwara et al., 2018; Spratt et al., 2019).

Utilizing Gabra6-Cre, Na$_v$1.2 has also been knocked down or deleted from cerebellar granule neurons (C. Wang et al., 2024). Unlike the cerebrum, where *Scn2a* gets downregulated in adulthood, cerebellar granule neurons express *Scn2a* at high levels throughout life (Gazina et al., 2010). When *Scn2a* was deleted in these cells using Gabra6-Cre, mice displayed a hypersensitive vestibular ocular reflex (VOR) gain (C. Wang et al., 2024). Interestingly, this parallels the VOR changes in humans with pathogenic *SCN2A* mutations (Carson et al., 2017; Sanders et al., 2018; C. Wang et al., 2024). The change in VOR is likely due to alterations between cerebellar granule neurons and Purkinje cell synapses. Heterozygous loss of Na$_v$1.2 is most likely reducing the high-frequency transmission between cerebellar granule neurons and Purkinje cells, which can disrupt synaptic plasticity by impairing long-term potentiation and thus altering VOR (C. Wang et al., 2024).

Although Na$_v$1.2 is generally studied in neurons, it also functions in other nervous system cells. Gould & Kim, 2021 demonstrated that Na$_v$1.2 is necessary for spiking activity and maturation of a subset of oligodendroglia in the brainstem and cerebellum (Gould & Kim, 2021). To discover this, a *Scn2a$^{fl/f}$/Actin-Cre* line was used to globally delete *Scn2a* in mice around P4 to P6 with tamoxifen injection (Gould & Kim, 2021). Deletion of Na$_v$1.2 removed the subpopulation of spiking oligodendroglia but did not impact global oligodendroglia differentiation and maturation (Gould & Kim, 2021). Gould & Kim, 2021 also confirmed the presence of these *Scn2a* expressing oligodendroglia in the white matter of Olive baboons, a non-human primate (Gould & Kim, 2021). These data demonstrate that Na$_v$1.2 plays an integral role in neuronal signaling and can impact the development of non-neuronal cells, which could explain the heterogeneous nature of the phenotypes seen in humans.

**Patient-derived *Scn2a* LOF Models**

Currently, only one *SCN2A* model primarily functions as a LOF mutation. This model emulates the human variant p.R102X, a premature stop codon at position 102 in the protein, which in humans has a dominant negative effect (Kamiya et al., 2004; Ogiwara et al., 2018).

*Scn2a* [R102X/+] mice were identified as having absence-like seizures although appearing outwardly healthy, similar to the Planells-Cases *Scn2a* [+/-] model. At least one *Scn2a* [R102X/+] mouse was recorded using ECoG–EMG having a convulsive seizure (Ogiwara et al., 2018). It would be interesting to know if the dominant negative effect identified in humans translates to the mouse model. There is also a naturally occurring *Scn2a* variant in the C3H mouse strain, p.V752F, that has a dominant negative effect (Oliva et al., 2014).

# Gain of Function *Scn2a* Mouse Models

### *Scn2a*$^{Q54}$

Shortly after the first *Scn2a* knockout model was published, Kearney et al. published the first *Scn2a* gain of function (GOF) model: the *Scn2a*$^{Q54}$ mouse. In this model, the neuron-specific enolase (NSE) promoter drives the expression of FLAG-tagged *Scn2a* GAL879-881QQQ, where the mutation is located in the evolutionarily conserved S4-S5 linker region of domain two (Kearney et al., 2001). These mice displayed spontaneous seizures that originated in the hippocampus and increased in number and duration with age, eventually leading to premature death around 4 to 9 months (Kearney et al., 2001; Kile et al., 2008). High-frequency stimulation of hippocampal neurons in *Scn2a*$^{Q54}$ mice leads to increased neuronal oscillations after discharge and spontaneous activity consistent with hyperexcitability and seizure activity (Kile et al., 2008). Examination of hippocampi from adult *Scn2a*$^{Q54}$ mice determined that the frequent and prolonged seizures caused significant neuronal loss and gliosis in the CA1, CA2, and CA3 regions of the hippocampus (Kearney et al., 2001; Kile et al., 2008). Furthermore, grooming, although preserved in *Scn2a*$^{Q54}$ mice, was rigid in structure and had atypical characteristics like prolonged bouts, abnormal postures, and repetitive motions (Kile et al., 2008). The rigid and irregular pattern of grooming in *Scn2a*$^{Q54}$ mice could be indicative of autism-like symptomatology and may further suggest changes in neural pathways that are influencing the microstructure of innate behaviors (H. Kim et al., 2016). A summary of *Scn2a*$^{Q54}$ findings is provided in Figure 5.

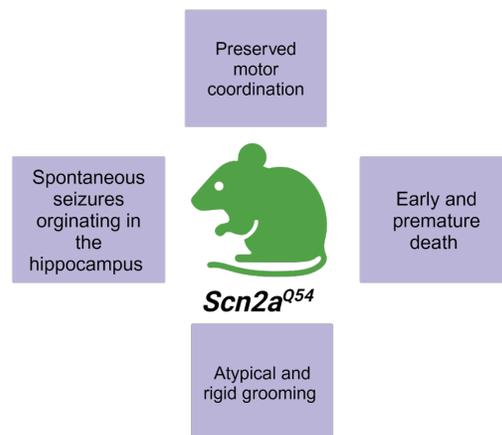

**Figure 5:** Summary of findings utilizing the *Scn2a*$^{Q54}$ mouse construct on the mixed C57BL/6J x SJl/J background. Created in BioRender. Williams lab, A. (2024) https://BioRender.com/h22y889

### Background Strain Effects in *Scn2a*$^{Q54}$ Mice

The *Scn2a*$^{Q54}$ mouse was originally engineered on a mixed C57BL/6J x SJl/J (F1.*Scn2a*$^{Q54}$) background and had a severe seizure phenotype. However, when crossed to a congenic C57BL/6J background (B6.*Scn2a*$^{Q54}$), the seizure phenotype attenuated (Bergren et al., 2005, 2009; Thompson et al., 2017). Specifically, F1.*Scn2a*$^{Q54}$ mice display both greater evoked activity and increased firing frequency within the hippocampus compared to the B6.*Scn2a*$^{Q54}$ mice, although B6.*Scn2a*$^{Q54}$ mice still display increased hippocampal activity compared to wildtype mice (Thompson et al., 2017). The level of persistent sodium current and depolarized inactivation of hippocampal pyramidal cells also correlated with the changes in firing frequency and evoked potentials between the two Q54 models (Thompson et al., 2017).

The change in seizure phenotype between background strains created an avenue to investigate genetic modifiers of *Scn2a*-related disorders. Identifying genetic modifiers of *SCN2A* could provide insight into why individuals with identical mutations do not present clinically with matching symptomology and severity. Genetic screening of both background strains identified loci on chromosomes 11 and 19, Modifier of epilepsy 1 (Moe1) and Modifier of epilepsy 2 (Moe2), respectively (Bergren et al., 2005). These loci accounted for around 80% of the observed differences in seizure susceptibility between the two strains (Bergren et al., 2005). Furthermore, the SJI background strain carries *doubleridge (dblr)*, a transgene-induced mouse mutation, in chromosome 19, which is also the location of the Moe2 locus (MacDonald et al., 2004). The *dblr* mutation reduces the expression of Dickkopf-1(*Dkk1*), a Wnt signaling inhibitor, and has been related to increased seizure events (MacDonald et al., 2004). However, reducing levels of *Dkk1* in B6.*Scn2a$^{Q54}$* mice by crossing them with *Dkk1$^{+/-}$* mice did not increase seizure susceptibility or incidence; thus, *Dkk1* is unlikely to be the genetic modifier of seizure severity between the two background strains of *Scn2a$^{Q54}$* mice (MacDonald et al., 2004).

Fine mapping of chromosome 19 determined that the Moe2 locus within the *dblr* region is 21.8kb upstream of *Dkk1* (Bergren et al., 2009), thus supporting previous data indicating that *Dkk1* was not the modifier increasing seizure susceptibility. However, careful chromosome mapping identified twenty-four other potential genetic modifiers within the Moe2 locus (Bergren et al., 2009). Sequencing of the candidate genes uncovered non-synonymous coding sequence polymorphisms in the potassium channel gene *Kcnv2* (voltage-gated potassium channel subfamily V, member 2/$K_v$8.2) and the transcription factor *Smarca2* (SWI/SNF related matrix-associated, actin-dependent regulator of chromatin, subfamily a, member 2) (Bergren et al., 2009). Ultimately, *Kcnv2* was determined to be the strongest functional candidate of Moe2 locus involvement in the seizure phenotype differences observed between the F1.*Scn2a$^{Q54}$* and B6.*Scn2a$^{Q54}$* mice (Bergren et al., 2009). Knowing how genetic modifiers influence the *Scn2a$^{Q54}$* mouse phenotype helps elucidate how similar mutations can lead to disparate seizure and symptom profiles of individuals with *SCN2A*-related disorders.

### *Scn2a$^{Q54}$* Mice and Genetic Modulators of Seizure Severity

The pore-forming subunit of the T-type calcium channel Ca$_v$3.1, *Cacna1g,* has been identified as a genetic modifier of seizure severity between the B6.*Scn2a$^{Q54}$* mice and the F1.*Scn2a$^{Q54}$* mice (Calhoun et al., 2016). *Cacna1g* is located in the Moe1 locus. Elevation of Ca$_v$3.1 correlated with increased spontaneous seizure activity, while a decrease in Ca$_v$3.1 correlated with decreased seizure activity (Calhoun et al., 2016). Cacna1g could be a potential therapeutic target for the treatment of *Scn2a*-related epilepsies that do not respond to sodium or potassium channel blockers.

Hepatic leukemia factor (*Hlf*) is another genetic modifier of seizures in *Scn2a$^{Q54}$* mice (Hawkins & Kearney, 2016). Deletion of *Hlf* in B6.*Scn2a$^{Q54}$* mice augments the seizure phenotype (Hawkins & Kearney, 2016). *Hlf* is a regulator of the pyridoxine pathway, and it was postulated that deletion of *Hlf* exacerbated the severity of seizures through dysregulation of the pathway. When fed a diet deficient in pyridoxine, the number and severity of seizures of the B6.*Scn2a$^{Q54}$* mice also increased (Hawkins & Kearney, 2016). These findings reveal pyridoxine as a regulator of seizure severity in this mouse model (Hawkins & Kearney, 2016).

Calmodulin protein kinase II (CaMKII), a modulator of persistent current and channel inactivation, alters and likely indicates seizure severity of *SCN2A*-related seizures (Thompson et al., 2017). Higher levels of CaMKII were found in hippocampal lysates from the F1.*Scn2a$^{Q54}$* mice (more severe phenotype) compared to B6.*Scn2a$^{Q54}$*mice (attenuated phenotype) (Thompson et al., 2017). Both *Scn2a$^{Q54}$* lines, however, had elevated levels of CaMKII present compared to wildtype, which indicates that CaMKII levels rise in response to increased Na$_v$1.2 activity. Lastly, inhibition of CaMKII in either F1.*Scn2a$^{Q54}$* or B6.*Scn2a$^{Q54}$* mice reduced neuronal excitability and persistent currents, improving the seizure phenotype (Thompson et al., 2017).

The identification of multiple modifiers of *Scn2a*-related seizures in mice can uncover potential new targets for the treatment of *SCN2A*-related seizure disorders.

**Other *Scn2a* GOF models**

Beyond *Scn2a*$^{Q54}$, there are at least two other GOF *Scn2a* mouse models, both of which model human variants. The first is the *Scn2a*$^{A263V/+}$ mouse, which displays increased activity in hippocampal CA1 pyramidal neurons (Schattling et al., 2016). The *Scn2a*$^{A263V/+}$ model was specifically developed to better understand sodium channel involvement in autoimmune encephalomyelitis (EAE) which leads to gradual disability through neuron loss (Schattling et al., 2016). The p.A263V mutant allele compounded the effect of EAE by increasing Na$_v$1.2 activity, leading to greater axonal injury and energy consumption (Schattling et al., 2016). These changes were noted in both sexes but were most prominent in male mice (Schattling et al., 2016). The p.A263V variant also exacerbated neurodegeneration associated with multiple sclerosis (Fazeli et al., 2016).

The other existing GOF model emulating a human variant is the *Scn2a*$^{Q/+}$ mouse (M. Li et al., 2021). The *Scn2a*$^{Q/+}$ mouse was generated by inserting the point mutation p.1883Q into exon 26 of the *Scn2a* gene. These mice displayed phenotypes consistent with developmental epileptic encephalopathy and were not viable after P30 due to frequent seizure activity (M. Li et al., 2021). Overall, *Scn2a* GOF mice have similar seizure phenotypes regardless of the specific mutation. This parallels patient data, as seizure severity is poorly predicted from simply identifying the pathogenic *SCN2A* variant (Berecki et al., 2022; Bergren et al., 2005, 2009; Crawford et al., 2021; Lauxmann et al., 2018; M. Li et al., 2021; Schattling et al., 2016).

## *Scn2a*-p.K1422E

*SCN2A*-p.K1422E is a pathogenic *de novo* variant of *SCN2A* located in the Na$_v$1.2 channel pore. This variant demonstrates both loss and gain of function characteristics in Nav1.2, which has made it difficult to classify (Echevarria-Cooper et al., 2022). P.K1422E impairs the ion selectivity filter of the Na$_v$1.2 channel, thus allowing other ions besides sodium to enter (Echevarria-Cooper et al., 2022). This mutation diminishes sodium conductance, which alters repolarization and neuronal firing (Echevarria-Cooper et al., 2022).

The mouse model emulating this *Scn2a* variant has been evaluated from both behavioral and electrophysiological perspectives. *Scn2a*$^{E/+}$ mice exhibit reduced anxiety-like behaviors compared to wildtype littermates on the elevated plus maze. However, on the open field test, anxiety-like behaviors are more pronounced in *Scn2a*$^{E/+}$ female mice compared to wildtype. *Scn2a*$^{E/+}$ mice also display hypersocial behaviors that may parallel what is seen in humans. These mice also do not have impairments in odor discrimination similar to other *Scn2a* haploinsufficient mice (Echevarria-Cooper et al., 2022; Léna & Mantegazza, 2019; Marcantonio et al., 2023). However, *Scn2a* mutant mice with extremely low levels of *Scn2a* expression or no *Scn2a* expression in ObGCs are slower to discriminate odors (Nunes & Kuner, 2018; Zhang et al., 2021). These findings indicate that there is a level of *Scn2a* that is required for accurate and efficient olfaction.

*Scn2a*$^{E/+}$ cortical neurons revealed slower action potential initiation, likely due to reduced sodium currents (Echevarria-Cooper et al., 2022). Similar characteristics have been identified in immature cultured cortical neurons from other *Scn2a* haploinsufficient models (Spratt et al., 2021). In vivo EEG recordings of *Scn2a*$^{E/+}$ mice displayed infrequent and spontaneous seizures localized to the posterior cortex. When seizures were induced with flurothyl, *Scn2a*$^{E/+}$ mice had a slower progression from myoclonic jerks to generalized tonic-clonic seizures (GTCS) compared to wildtype C57BL/6J littermates. Furthermore, the p.K1422E mutation was found to disrupt the estrous cycle in female mice but did not significantly alter the distribution of GTCS onset in female mice (Echevarria-Cooper & Kearney, 2023).

Similar to some GOF models, the *Scn2a*$^{E/+}$ mice demonstrate strain-dependent neurobehavioral and seizure phenotypes (Echevarria-Cooper et al., 2023). The mouse line was originally generated on the C57BL/6J (B6.*Scn2a*$^{E/+}$) background but was later crossed with DBA/2J (D2) mice to create a mixed background of [D2xB6], resulting in F1D2.*Scn2a*$^{E/+}$ animals (Echevarria-Cooper et al., 2022, 2023). These studies revealed that anxiety-like behaviors are more pronounced in the B6.*Scn2a*$^{E/+}$ mice (Echevarria-Cooper et al., 2023). F1D2.*Scn2a*$^{E/+}$ mice had a seizure profile similar to that of the original B6.*Scn2a*$^{E/+}$, however, a small subset displayed erratic behavior and then behavioral arrest (Echevarria-Cooper et al., 2023). Both B6.*Scn2a*$^{E/+}$ and F1D2.*Scn2a*$^{E/+}$ resisted seizure generalization with kainic acid, but the resistance was more pronounced in the F1D2.*Scn2a*$^{E/+}$ mice (Echevarria-Cooper et al., 2023). Furthermore, zero of the F1D2.*Scn2a*$^{E/+}$ mice reached stage 6 of the Racine scale (Echevarria-Cooper et al., 2022, 2023). Results from the *Scn2a*$^{E/+}$ mice highlight how *Scn2a* variants can present with characteristics of both gain and loss of function, and *SCN2A*-related disorders are likely much more complicated than the variants' channel kinetics.

## Scn2a Mice and Potassium Channels

The voltage-gated potassium channel, $K_v7.2$, has been shown to contribute to seizure phenotypes seen in *Scn2a*$^{Q54}$ mice (Kearney et al., 2006). This potassium channel generates M currents that modulate hippocampal neurons' firing pattern and excitability (Kearney et al., 2006). When Scn2a$^{Q54}$ mice also have a mutation in Kv7.2, which reduces the function and activity of the potassium channel, seizure severity of the *Scn2a*$^{Q54}$ mice is exacerbated and premature death occurs around P21 (Kearney et al., 2006). The intense seizure phenotype of the double mutant mice suggests that M currents play an important and likely necessary role in modulating seizure initiation and spreading (Aiba & Noebels, 2021). Knowing that M currents can regulate seizure severity in an *Scn2a* mouse model opens the possibility of developing a treatment targeting $K_v7.2$ to alleviate *SCN2A*-related seizures.

Potassium channels also modulate *Scn2a*$^{+/-}$ phenotypes (Indumathy et al., 2021; Mishra et al., 2017). Mice heterozygous for both *Scn2a* and *Kcn1,* a gene encoding for the pore-forming subunit of the voltage-gated potassium channel $K_V1.1$ *(Scn2a*$^{+/-}$;*Kcna1*$^{+/-}$), have reduced anxiety and autism-like behaviors compared to *Scn2a*$^{+/-}$ mice (Indumathy et al., 2021). In addition, the downregulation of potassium channels was found to contribute to the hyperexcitability of medium spiny neurons cultured from *Scn2a*$^{gtKO/gtKO}$ mice (Zhang et al., 2021). When pimaric acid, a global potassium channel agonist, was applied to *Scn2a*$^{gtKO/gtKO}$ medium spiny neurons, their excitability was largely normalized to that of wildtype animals (Zhang et al., 2021). Furthermore, the application of 4-trifluoromethyl-L-phenylglycine, a specific $K_v1.1$ agonist, improved the hyperexcitability phenotype of *Scn2a*$^{gtKO/gtKO}$ medium spiny neurons, which indicates that $K_v1.1$ may be a potential target for the treatment of some *SCN2A* related seizures (Zhang et al., 2021). Additionally, *Scn2a* deletion improved the survival of *Kcna1*$^{-/-}$ mice (Mishra et al., 2017). Lastly, RNA sequencing data from *Scn2a*$^{gtKO/gtKO}$ whole mouse brains indicate that multiple voltage-gated potassium channels are downregulated in response to the loss of $Na_v1.2$ (Zhang et al., 2021). Taken together, these data indicate that $Na_v1.2$ and voltage-gated potassium channels likely have a dynamic relationship that may prove useful from a therapeutic perspective.

As evidenced by the growing literature of *Scn2a* mouse models, voltage-gated potassium channels are likely important in *SCN2A*-related disorders. Thus far, $K_v1.1$, $K_v1.2$, $K_v7.2$, $K_v8.2$, and $K_v9.2$ have all been implicated in modulating $Na_v1.2$ channel activity and/or expression (Bergren et al., 2009; Indumathy et al., 2021; Kearney et al., 2006; Mishra et al., 2017; Nadella et al., 2022; Zhang et al., 2021). Potassium channel modulation should continue to be explored as a therapeutic target for the seizure treatment of individuals with pathogenic *SCN2A* variants (Berecki et al., 2022; Lauxmann et al., 2018). However, it is important to note that the decision to block or activate specific potassium channels to attenuate seizures would likely depend on each patient's specific variant.

## Treatment of *Scn2a* Related Phenotypes in Mice

**Genetic Rescue**

There have been multiple attempts to develop a viable and safe genetic therapy to alleviate *SCN2A*-related disorders. One such attempt explored *cis*-regulation therapy (CRT) by using a dCas9 fused to a transcriptional co-activator (Tamura et al., 2022). This method allows CRISPR activation (CRISPRa) to increase the expression of the functional wildtype *Scn2a* allele in both *Scn2a*$^{+/-}$ and *Scn2a*$^{+/kl}$ mice (Supplemental Table 1) (Tamura et al., 2022). The rAAV-CRISPRa-based treatment restored typical excitability patterns in *Scn2a*$^{+/-}$ cortical neurons compared to wildtype (Tamura et al., 2022). Furthermore, the treatment was not shown to induce seizures or lead to phenotypic abnormalities in the animals (Tamura et al., 2022). CRISPRa rescue and upregulation of *Scn2a* was then utilized to recover an appropriate vestibular ocular reflex (VOR) in *Scn2a*$^{+/-}$ mice which is known to be impaired in both humans with pathogenic variants of *SCN2A* and *Scn2a*$^{+/-}$ mice (C. Wang et al., 2024). Based on the promising results from the mouse models, it appears that using CRISPRa therapy could be a viable and safe avenue for treating *SCN2A* loss-of-function variants. Further data indicating the safety and efficacy of the CRISPRa treatment will be needed to truly know if this is an avenue of treatment that will benefit individuals like it appears to benefit the *Scn2a*$^{+/-}$ mice.

Another genetic therapy approach involves gapmer antisense oligonucleotides (ASO), which target *Scn2a* mRNA (M. Li et al., 2021). The ASO was developed to alleviate seizures experienced by *Scn2a*$^{Q/+}$ mice (M. Li et al., 2021). Two doses of the ASO ED were tested: ASO ED50 and ED80, which reduced *Scn2a* mRNA by 50% and 80%, respectively. ASO ED80 was found to be toxic in wildtype C57BL/6N mice, although it did extend the lifespan of *Scn2a*$^{Q/+}$ mice and attenuated seizure severity when utilized at P1 (M. Li et al., 2021). On the other hand, ASO ED50, when used at P1, did not impair weight, sociability, or motor behavior in wildtype C57BL/6N mice or *Scn2a*$^{Q/+}$ mice (M. Li et al., 2021). However, ASO ED50 did reduce anxiety-like behaviors in *Scn2a*$^{Q/+}$ mice below what was seen in treated wildtype C57BL/6N mice (M. Li et al., 2021). Treating *Scn2a*$^{Q/+}$ mice older than P1 with ASO ED50 still attenuated seizure severity and extended lifespan compared to untreated animals, which suggests some degree of flexibility in the treatment window (M. Li et al., 2021). From the reported data, it seems plausible that ASO therapy may benefit some *SCN2A* variants. However, ASO therapies need to be tailored to specific variants to be effective, and the heterogeneous nature of *SCN2A* reduces the possibility of generating a large-scale treatment of *SCN2A*-related disorders utilizing this approach (Lauffer et al., 2024). Furthermore, Tofersen, the only FDA-approved ASO, is associated with many adverse side effects and has demonstrated limited therapeutic benefit (Lauffer et al., 2024). Treatments utilized in *SCN2A*-related disorders must be safe and effective as it is likely to be provided to children, and aversive side effects may irreversibly alter development.

**Drug Treatment**

Several pharmacotherapies have been tried in *Scn2a* mouse models to attenuate symptoms associated with pathogenic variants. One such drug is CX516, a positive allosteric modulator of AMPA receptors (Tatsukawa et al., 2019). Intraperitoneal injection of CX516 at least 10 minutes prior to behavioral tasks at multiple doses was shown to reduce hyperactivity phenotypes seen in *Scn2a*$^{+/-}$ mice (Tatsukawa et al., 2019). The greatest effect was seen at the highest dose, 40 mg/kg, and the treatment did not appear to negatively affect wildtype C57BL/6J mice (Tatsukawa et al., 2019).

Other drugs tested in *Scn2a* mutant mice have been utilized to attenuate seizures. These drugs include Ranolazine and GS967/ PRAX-562, which both inhibit persistent sodium current (Anderson et al., 2014). Both of these drugs reduced seizure frequency in F1.*Scn2a*$^{Q54}$ mice with GS967/ PRAX-562 being more effective (Anderson et al., 2014). GS967/ PRAX-562 further improved survival of F1.*Scn2a*$^{Q54}$ mice and prevented hilar neuron loss and protected the mice from induced seizures in the maximal electroshock paradigm (Anderson et al., 2014). However, GS967/ PRAX-562 suppressed hippocampal mossy fiber sprouting in these mice, which could lead to the alteration of behavioral phenotypes, like spatial learning (Comba et al., 2015; Ramírez-Amaya et al., 2001). Further data from mice treated with GS967/ PRAX-562 are needed to evaluate potential adverse effects. Although these sodium channel blockers are effective in treating the GOF seizure phenotype of F1.*Scn2a*$^{Q54}$ mice, they would likely be an ineffective, if not harmful, treatment of seizures for individuals with LOF *SCN2A* variants (H. J. Kim et al., 2020; Miao et al., 2020; Oyrer et al., 2018; Reynolds et al., 2020; Thompson et al., 2017).

## Other Rodent Models of *Scn2a*

Research performed in rats established our foundational understanding of *SCN2A*, like the temporal patterning of *SCN2A* expression (Gong et al., 1999) and the subcellular localization of Na$_v$1.2 (Westenbroek et al., 1989). In Sprague Dawley rats, Na$_v$1.2 protein is expressed within the forebrain, substantia nigra, hippocampus, and the molecular and granular layers of the cerebellum (Westenbroek et al., 1989) while Na$_v$1.2 mRNA was highly expressed within the pyramidal and granule layer of the dentate gyrus as well as the granule layer of the cerebellum (Black et al., 1994). Na$_v$1.2 was further found to have a complementary expression pattern with Na$_v$1.1 throughout the rat cortex and preferentially localized to axons (Gong et al., 1999). Further background work utilizing Sprague Dawley rats revealed a population of Na$_v$1.2 lacking the beta-2 subunit during the first two weeks of life that was not present in older animals, indicating that *Scn2a* undergoes developmental stage-specific changes (Gong et al., 1999). Recently, a Long Evans rat *Scn2a*$^{+/-}$ model was engineered using CRISPR-Cas9 to disrupt exon 5 of *Scn2a* (Kastner et al., 2024), which generates a frameshift and presumed LOF. On a novel choice-wide behavioral association test, the *Scn2a*$^{+/-}$ rats displayed impairments in spatial alternation-related behaviors compared to wildtype controls (Kastner et al., 2024).

Currently, an *Scn2a*$^{+/-}$ prairie vole is being characterized (Perry Spratt 2020). Prairie voles have social structures more similar to humans than rats or mice and may better emulate the social changes seen in human *SCN2A* disorders. The *Scn2a*$^{+/-}$ prairie vole model is being engineered using CRISPR-Cas9 mutagenesis to introduce a premature stop codon within the first and third coding exons of the *Scn2a* gene (Perry Spratt 2020). Preliminary data from the early generations of the *Scn2a*$^{+/-}$ prairie voles indicate impaired dendritic excitability similar to the *Scn2a*$^{+/-}$ mouse models (Perry Spratt 2020). This may be the first evidence that dendritic deficits caused by loss of *Scn2a* are conserved across species.

## Discussion

**Why are there Differential Findings in *Scn2a* Rodent Models?**
Loss of Na$_v$1.2 leads to many changes in neuronal activity that can be specific to cell type and developmental stage. Severe loss of Na$_v$1.2 can lead to paradoxical increases in the excitability of mature cortical-striatal and neocortical pyramidal neurons while also impairing dendritic excitability (Miyamoto et al., 2019; Spratt et al., 2021; Zhang et al., 2021). Nevertheless, loss of Na$_v$1.2 in immature *Scn2a*$^{+/-}$ pyramidal neurons leads to reduced excitability similar to the effects seen in obGCs and somatosensory cortical neurons (M. Li et al., 2023; Nunes & Kuner, 2018; Spratt et al., 2019). Since distinct brain regions and cell types differentially contribute to many behaviors, and the functional impact of Na$_v$1.2 loss differs based on neuronal cell type and developmental stage, this could be a source of apparently conflicting findings in *SCN2A* models. When conducting behavioral assays in *Scn2a* mutant mice, it is important to consider what cell type or brain region is involved in the behavior being assessed since the functional role of *SCN2A* may be variable across paradigms. It is also paramount that when electrophysical consequences of Na$_v$1.2 loss are investigated, the developmental stage and cell type are reported, as that information can drastically change interpretation of results.

Another factor that could contribute to the inconsistent behavioral findings is the age of the mice being tested. At least one study to date found that behavioral phenotypes attenuate as *Scn2a*$^{+/-}$ mice age, and multiple studies have identified electrophysiological alterations that are only present during the first postnatal week (Léna & Mantegazza, 2019; Ogiwara et al., 2018; Spratt et al., 2021). Na$_v$1.2 is known to play a significant role in adult animal's excitatory neurons but may also influence inhibitory neurons early in development (Miyamoto et al., 2019). Moreover, there are two isoforms of Na$_v$1.2: the neonatal and adult. The neonatal isoform of Na$_v$1.2 is less excitable than the adult isoform and is replaced by the adult isoform starting around P9 in mice (Gazina et al., 2010; Gazina et al., 2015). Some *SCN2A* variants exhibit greater and more severe dysfunction in the neonatal isoform than the adult (Muller, 2020; Thompson et al., 2020; Ben-Shalom et al., 2017). For example, *SCN2A* variants associated with SeLFNIE exhibit early-onset seizures that remit within the first few years of life (Xu et al., 2007). These findings support the idea that the two isoforms of *SCN2A* have different roles in neuronal excitability and could differentially contribute to observed phenotypes. Therefore, it is reasonable to suggest that variation in the ages of the *Scn2a* mutant mice evaluated for behavior may contribute to some of the differential findings. To determine the full developmental and functional impact of Na$_v$1.2 variants, regardless of isoform, *SCN2A* models should be explored from a developmental perspective, and results should be reported with those details.

It is key to note that *SCN2A* is not a sex-linked gene, and therefore disorders secondary to *SCN2A* mutations have an equal likelihood of occurring in both males and females at similar rates. However, *SCN2A* variants could interact with sex-dependent genes and hormones differently. For instance, the p.K1422E variant was found to disrupt the estrous cycle of female mice, which could lead to different behavioral phenotypes between male and female mice (Echevarria-Cooper & Kearney, 2023). Without accounting for hormonal cycles, multiple investigations of *Scn2a* mice have already noted behavioral differences between male and female mice (Echevarria-Cooper et al., 2022; Marcantonio et al., 2023; Schattling et al., 2016; Spratt et al., 2019, 2021). Thus, it is paramount that *SCN2A*-related disorders continue to be equally investigated in male and female animals, as sex differences could create differences in the presentation of symptoms.

Lastly, the background strain of the animals themselves likely contributes to the contrasting behavioral results between similar *Scn2a* models. In work with GOF *Scn2a* models the genetic background strain has been shown to alter the severity of displayed phenotypes (Bergren et al., 2005; Thompson et al., 2017). Experiments completed with different genetic strains highlight the importance of considering both genetic and epigenetic modifiers when thinking about how *Scn2a* variants lead to disease. It has yet to be determined which background strain most accurately and reliably recapitulates human symptomatology of *Scn2a*-related disorders.

**Gain and Loss of Function Classification of *SCN2A* mutations; is it Enough?**

*SCN2A*-related disorders can drastically change a child's quality of life; therefore, it is imperative to determine the varying pathologies and prognoses enumerated by *SCN2A* variants. Currently, it is thought that *SCN2A*-related disorders bifurcate into two categories: neurodevelopmental disorders caused by LOF or seizure disorders caused by GOF in Na$_v$1.2. GOF mutations often result in early-onset seizure disorders (Berecki et al., 2022; Lauxmann et al., 2018), while LOF mutations are linked to diagnoses of Autism Spectrum Disorder (ASD) and/or intellectual disability (ID) and other neuropsychiatric conditions (Ben-Shalom et al., 2017). While this binary organization has simplified the classification of Na$_v$1.2 channel function, it does not fully capture the nuances of *SCN2A*-related disorders.

Although LOF variants are largely associated with ASD and ID, an estimated 30% of individuals with LOF mutations develop severe seizure disorders that are often intractable (Reynolds et al., 2020; Thompson et al., 2020). Data from LOF *SCN2A* models indicate that compensatory and maladaptive changes in potassium channels can lead to the onset of seizures (Spratt et al., 2021; Zhang et al., 2021). There are also LOF variants that lead to the development of seizures prior to 3 months of age in humans, including p.R853Q, p.G899S, and p.P1658S (Berecki et al., 2018; Miao et al., 2020; Wolff et al., 2017). As the age of seizure onset does not always align with the true electrophysical profile of the channel variant, (Berecki et al., 2022)) argued that it should not be the sole classifier of an *SCN2A* channel variant. Furthermore, misassignment of the GOF or LOF label to a *SCN2A* channel variant can be detrimental as treatment of a LOF variant with sodium channel blockers can exacerbate seizures and lead to unfavorable outcomes for the affected individual (Miao et al., 2020; Oyrer et al., 2018, p. 8; Reynolds et al., 2020; Thompson et al., 2020). Additionally, the age of onset and the electrophysical properties of *SCN2A* GOF channel variants are poor predictors of severity of seizures, suggesting we should seek better biomarkers of disease progression and severity (Berecki et al., 2022; Crawford et al., 2021; Lauxmann et al., 2018).

Human data also indicate that identical channel variants can lead to different phenotypes and diagnoses across individuals (Berecki et al., 2022; Passi & Mohammad, 2021). For instance, the autosomal dominant *SCN2A* variant p.L1650P presented in a male child as episodic ataxia but in the father as episodic hemiplegia, two very distinct diagnoses (Passi & Mohammad, 2021). Similarly, the variants p.R1319Q and p.V261M lead to severe developmental delay in some individuals, while others with the variants have typical development (Zeng et al., 2022). Likewise, the variants p.R102X and p.R1435X led to seizures in some individuals but no seizures in others (Begemann et al., 2019; Kamiya et al., 2004; Trump et al., 2016; Wolff et al., 2017). Currently, it is unclear what drives the differences between individuals with identical channel variants, but it is clear that the consequences of channel variants are not fully explained by the variants themselves. Patients with similar or identical pathogenic variants should undergo whole genome and/or epigenome sequencing as their unique genetic and epigenetic profiles could explain at least in part their dissimilar symptomology. Data from *Scn2a* mouse models support the idea that genetic modifiers can drive symptom presentation and severity, but genetic modifiers have not been explored in humans (Bergren et al., 2005, 2009; Calhoun et al., 2016; Hawkins & Kearney, 2016; MacDonald et al., 2004; Passi & Mohammad, 2021; Thompson et al., 2017).

There are also now reports of *SCN2A* variants displaying both GOF and LOF characteristics, defying categorization by the binary system. For instance, the p.K1422E variant located on the channel pore of Na$_v$1.2 displays mixed biophysical characteristics to the extent that the authors who originally characterized the variant were not comfortable assigning it a classification (Echevarria-Cooper et al., 2022). Another variant, p.G879R, was found to have a largely LOF profile but with an increased degree of overlap between the activation and inactivation of sodium channel currents typical of a GOF channel variant (Yang et al., 2022). The biophysical profile of a channel variant can also shift depending on the *SCN2A* isoform present, like in the case of p.R1882L and p.R1882Q (Thompson et al., 2023). Both variants display GOF characteristics while neonatal Na$_v$1.2 is expressed but display mixed biophysical properties when the adult isoform replaces the neonatal (Thompson et al., 2023). Data from individuals with *SCN2A*-related disorders, SeLFNIE (previously BFNIS), further highlight the influence that *SCN2A* isoforms have on pathogenic variant symptom presentation (Xu et al., 2007). These individuals usually experience typical development and display early-onset seizures that attenuate with age (Xu et al., 2007). From non-human mammalian models, we can garner that this is likely due to the different electrophysical properties of the neonatal and adult channel isoforms (Gazina et al., 2015). The pathogenic *SCN2A* variants leading to SeLFNIE likely cause the neonatal Na$_v$1.2 isoform to behave more like the adult isoform. Therefore, when the adult isoform replaces the neonatal, the biophysical properties of the Na$_v$1.2 channel likely better match the developmental state of the individual and attenuate the seizure phenotype. These data demonstrate the need to study the developmental effects and trajectory of the *SCN2A* gene and its variants.

A recent study utilized human phenotype ontology terms to evaluate potential correlations between diagnostic terms and *SCN2A* variants (Crawford et al., 2021). The analysis revealed that missense *SCN2A* variants are largely correlated with terms like "neonatal onset," "seizures,' and "no intellectual disability", while premature termination variants were correlated with terms like "behavioral abnormality," "autism," "autistic behavior," and " no seizures" (Crawford et al., 2021). Although ASD and ID diagnoses are primarily associated with premature termination variants, these diagnoses can still be related to missense *SCN2A* variants. Furthermore, if seizures, developmental delay (DD), or intellectual disability (ID) are profound, it can be impossible to disentangle symptoms of autism spectrum disorder (ASD) from existing comorbidities, so the diagnosis is not made (Srivastava & Sahin, 2017). There is at least one reported variant, p.L1342P, where the patient presented with ID and early seizure onset but also displayed no eye contact, which is a common trait of ASD (Begemann et al., 2019). It is possible that the comorbidities of existing diagnoses masked a potential diagnosis of ASD (Srivastava & Sahin, 2017).

The phrase "When you hear hoof beats, don't look for zebras" is often used in medicine to encourage providers to look for common diagnoses, but *SCN2A*-related disorders are not common; each variant produces a unique profile of symptoms. Understanding whether a variant is gain-of-function (GOF) or loss-of-function (LOF) can guide initial treatment, but it is unlikely to capture the complete biophysical and symptom profile of every individual. *SCN2A*-related disorders do not occur within a binary but a spectrum of diagnoses, and the biophysical characteristics of channel variants are likely just as complex.

Acknowledgements: This work was supported by iDREAM (R25 NS130966) to MFHA, the Neuroscience Graduate Program T32 (T32 NS007421) (KS), and the Department of Defense Autism Research Program (DOD AR220030) (AJW).

| Model | Principle Paper | Related Papers | Iteration of Conditional Model |
|---|---|---|---|
| **Constitutive Haploinsufficient Mouse Models of SCN2A** | | | |
| *Scn2a +/-* | Planells-Cases et al., 2000 | Planells-Cases et al., 2000 | |
| | | Mishra et al., 2017 | |
| | | Middleton et al., 2018 | |
| | | Ogiwara et al., 2018 | |
| | | Lena and Mantegazza 2019 | |
| | | Spratt et al., 2019 | |
| | | Tatsukawa et al., 2019 | |
| | | Miyamoto et al., 2019 | |
| | | Spratt et al., 2021 | |
| | | Indumathy et al., 2021 | |
| | | Tamura et al., 2022 | |
| | | Schamiloglu et al., 2023 | |
| | | Wang et al., 2024 | |
| | | Marcantonio et al., 2023 | |
| *Scn2a+/–; Kcna1-/-* | Mishra et al., 2017 | Mishra et al., 2008 | |
| | | Indumathy et al., 2021 | |
| *Scn2a+/–; Kcna1+/–* | | Indumathy et al., 2021 | |
| *NaV1.2 adult* | | Gazina et al., 2015 | |
| *Scn2a fl/+* | | Shin et al., 2019 | |
| *Scn2a +/KI* | Tamura et al., 2022 | Tamura et al., 2022 | |
| **Premature Stop Codon Models** | | | |
| *Scn2a R102X/+* | | Ogiwara et al., 2018 | |
| **Conditional Mouse Models of Scn2a Haploinsufficiency** | | | |
| *Ei14::Scn2a+/-* | Spratt et al., 2019 | Spratt et al., 2019 | PV-Cre |
| | | | SOM-Cre |
| | | | CaMKIIα-Cre |
| *Scn2a +/fl* | | Spratt et al., 2021 | EF1α-Cre (Scn2a +/fl and Scn2a fl/fl) |
| | | Wang et al., 2024 | Alpha6-Cre |
| *SCN2A f/f* | Gould and Kim 2021 | | PDGFRA - Cre |
| | | | Actin- Cre |
| *Scn2a fl/+* | Ogiwara et al., 2018 | Yamagata et al., 2017 | Vgat-Venus |
| | | Ogiwara et al., 2018 | Emx1-Cre |
| | | | Vgat-Cre |
| | | | Vgat-Venus |
| | | Tatsukawa et al., 2019 | Emx1-Cre |
| | | | Vgat-Cre |
| | | Miyamoto et al., 2019 | Emx1-Cre |
| | | | Vgat-Cre |
| *Scn2a fl/fl* | | Miyamoto et al., 2019 | Trpc-Cre |
| | | | Ntsr-Cre |
| | | Suzuki et al., 2023 | mPFC-Cre |
| | | | VTA-Cre |
| *Scn2a +/KI* | Tamura et al., 2022 | | EF1α-Cre |
| *Scn2a fl/fl* | Ma et al., 2022 | | SCN-Cre |
| **Additional Mouse Models of Scn2a Reduction** | | | |
| *Scn2a :OB GC - shRNA* | | Nunes and Kuner, 2018 | |
| *Scn2aΔ1898/+* | | Wang et al., 2021 | |
| *Scn2a gtKO/gtKO* | Eaton et al., 2021 | Eaton et al., 2021 | |
| | | Ma et al., 2022 | |
| | | Wu et al., 2024 | |
| *Scn2a ASO* | | Li et al., 2023 | |
| **SCN2A Gain of Function Mouse Models** | | | |
| *Scn2a Q54* | Kearney et al., 2001 | Kearney et al., 2001 | |
| | | MacDonald et al., 2004 | |
| | | Bergren et al., 2005 | |
| | | Bergren et al., 2009 | |
| | | Kile et al., 2008 | |
| | | Anderson et al., 2014 | |
| | | Kim et al., 2016 | |
| | | Calhoun et al., 2016 | |
| | | Hawkins & Kearney, 2016 | |
| | | Thompson et al., 2017 | |
| *Scn2a A263v /+* | | Schattling et al., 2019 | |
| *Scn2a Q/+* | | Li et al, 2021 | |
| **Other SCN2A Mouse Models** | | | |
| *Scn2a E/+* | Echeveria-Cooper et al., 2022 | Echeveria- Cooper et al., 2022 | |
| | | Echeveria- Cooper et al., 2023 | |
| | | Echeveria- Cooper and Keanrey, 2023 | |

| Paper | Model | Iteration of Conditional Model | Mice Age | Social | Learning/Memory | Motor | Repetitive Behaviors | Anxiety/Hyperactivity | Olfaction | Seizures | Other | |
|---|---|---|---|---|---|---|---|---|---|---|---|---|
| | | | | | | | **Planells-Cases *Scn2a* +/- Mouse Model** | | | | | |
| Mishra et al., 2017 | | | 2 - 3 months | | | | | | | -ND in seizure threshold or epileptiform activity -Tendency to myoclonic jerk -ND in latency to GTC seizures | -ND in cardiac activity with EEG-ECG recordings -ND in brain-heart association | **RED** = Decrese **BLUE**= Increase/altered ND= No Difference |
| Middleton et al., 2018 **(males only)** | | | 4 - 6 months | | -Delayed learning in spatial working memory task -Slower learning in Barnes maze | | | | | | -Decrease in the peak firing rate of pyramidal cells during SPW-R events -Lower strength of assembly during SPW-Rs -Truncated hippocampal replay sequences | **OFT** = Open Field Test **EPM** = Elevated Plus Maze **FST** = Forced Swim Test **TST** = Tail Suspension Test **USV** = Ultrasonic Vocalization **PPI** = Prepulse Inhibition **mPFC** = Medial Prefrontal Cortex **ECoG** = Electrocorticogram **PTZ** = Pentylenetetrazol **FSI** = fast-spiking interneurons **EPSCs** = excitatory postsynaptic currents **VOR** = Vestibular Ocular Reflex |
| Ogiwara et al., 2018 | *Scn2a* +/- (Planells-Cases model) | | 6 - 11 weeks | | | | | | | -ND in susceptibility to hyperthermia induced seizures -Frequently abnormal ECoG recordings with absence-like SWDs -Higher maximum amplitude and spike numbers during a SWD episode -Shorter latency to have absence seizure-like immobility when injected with PTZ | -Higher prevalence and incidence of epileptiform discharges in mPFC -Impaired excitatory neural activity | |
| Lena and Mantegazza, 2019 **(males only)** | | | Juvenile: P22 - P44 | -Decreased amount and length of USVs produced -Decreased social interaction in the first two minutes of reciprocal social interaction test | -Slight decrease of spontaneous alternations in Y maze -Decreased preference for novel object in NOR task | -Increased acquired motor learning on rotarod | -Increased grooming -Increased marble burying | -Reduced anxiety in EPM -Reduced anxiety in OFT -ND in locomotion in OFT | -ND in odor discrimination task | | -Decreased immobility in TST | |
| | | | Adult: P60 - P95 | -Decreased amount and length of USVs produced -ND in reciprocal social interaction test | -ND in number of visits or spontaneous alternations in Y maze -ND in preference for novel object in NOR task -Slightly slower acquisition latency in Barnes maze test | -ND in motor learning in rotarod | -ND in grooming -ND in marble burying | -ND in anxiety in EPM -ND in anxiety in OFT -ND in locomotion in OFT | -ND in odor discrimination task | | -Slight decrease in immobility in minutes four to six of TST | |
| Spratt et al., 2019 | | | Age not reported | -Females showed no preference for novel stimuli in two-chamber social task | -Males had slightly impaired reversal learning in water T-maze -ND in spontaneous alternation assay | -ND in rotarod -ND in balance beam | -ND in grooming -ND in nesting assay | -Females had reduced anxiety in EPM -ND in OFT | | | | |
| Tatsuwaka et al., 2019 **(males only)** | | | P6 for USV P8 - 31 weeks for other assays | -Increased sociability in three-chamber assay -Slight dominant behavior in tube test -Deficit in social approach toward stranger mice in resident-intruder test -ND in social communication in USV | -Enhanced fear conditioning retention -Impaired extinction of fear-related memory | -Deficient motor coordination in rotarod | -Increased rearing in OFT | -Reduced anxiety in EPM -Activity in EPM -Reduced anxiety in OFT -Activity in OFT -Increased anxiety in light/dark assay | | | -Higher immobility in FST -Higher immobility in TST -ND in PPI -Increased activity in the gamma band in the mPFC | |
| Miyamoto et al., 2019 | | | > 8 weeks | | | | | | | -Have SWDs during behavioral quiescence | -Slight decrease in EPSCs in FSIs | |
| Indumathy et al., 2021 | | | 28 - 34 days | -ND in sociability in three-chamber assay -Decreased preference for novel stranger mice in social novelty assay | | | -Reduced marble burying -Increased grooming | | | | -ND in nestlet shredding | |
| Schamiloglu et al., 2023 | | | P54 - P119 | | -ND in learning the standard dynamic foraging task or variations of the assay where mice are head-fixed, the block length varies, there is reward optimization, or the probabilities vary -Slight increase in frequency to pick higher reward side in learning assay -Males had inflexible choice behavior during mid-block performance in the unpredictable paradigm of the dynamic foraging assay -ND in intertrial interval in dynamic foraging assay | | | | | -Spontaneous behavioral seizures in four of twenty five mice at the end of a day's training and one of the four mice died | | |
| Wang et al., 2024 | | | P60 - P90 | | | | | | | | -Consistently elevated VOR gain in different rotation frequencies -Decreased Purkinje cell firing rate | |
| | | | Juvenile: P21 - P44 | -Higher sociability in three-chamber assay -ND in social novelty in three-chamber assay | -ND in preference for displaced object in novel object recognition task -Lower entries into the arms of the Y maze -Increased decision-making latency in Y maze | -Higher latency to fall on day one of rotarod | -ND in grooming -ND in marble burying | -ND in EPM locomotion -Reduced anxiety in EPM -ND in anxiety in OFT -ND in OFT locomotion | -ND in odor discrimination task | | -ND in latency of reaction to stimuli in hotplate assay -ND in latency before withdrawal in coldplate assay | |

| Paper | Model | Iteration of Conditional Model | Mice Age | Social | Learning/Memory | Motor | Repetitive Behaviors | Anxiety/Hyperactivity | Olfaction | Seizures | Other |
|---|---|---|---|---|---|---|---|---|---|---|---|
| Marcantonio et al., 2023 (females only) | | | Adult: >P60 | -Higher sociability in three-chamber assay -Altered interaction with novel mouse in three-chamber assay | -ND in preference for object in NOR task -Lower entries into the three arms of Y maze -Increased decision-making latency in Y maze -Reduced percent of spontaneous alternations in Y maze | -ND in latency to fall in rotarod | -ND in grooming -ND in marble burying | -ND in EPM locomotion -Reduced anxiety in EPM -ND in anxiety in OFT -ND in OFT locomotion | -ND in odor discrimination task | | -ND in latency of reaction to stimuli in hotplate assay -Decreased latency before withdrawal in coldplate |
| **Other Constitutive Haploinsufficient Mouse Models Behavior** ||||||||||||
| Mishra et al., 2017 | Scn2a+/-; Kcna1-/- | | 2-3 months | | | | -ND grooming -Reduced nestlet shredding and marble burying | | | Scn2a+/– decreases spontaneous seizure duration in Kcna1–/– mice | Nav1.2 reduction does not ameliorate cardiac phentoypes in Kcna1–/– mice. |
| Indumathy et al., 2021 | Scn2a+/–; Kcna1+/– | | 28-34 days | ND in sociablity or social novelty | | | -ND in grooming and nestlet shredding -Reduced marble burying | | | | |
| Gazina et al., 2015 (males only) | NaV1.2 adult | | 53 days >= | -ND in social interaction -16x more likely to jump out of social interaction box | ND in memory and spatial learning on Y-maze and Morris water maze | ND on rotarod, locomotion or in grip strength | | -ND in OFT or Light-Dark box -Reduced anxiety in EPM | | Spontaneous seizures -More susceptible to PTZ-induced seizures | ND in thermal pain response, acoustic startle, PPI, or weight |
| Shin et al., 2019 (males only) | Scn2a f/+ | | 2-4 months | -ND social approach and communication -Increased direct social interation | -Reduced Spatial learning and memory on Morris water maze -Increased Fear memory | ND on Rotarod -ND in locatmotion in novel enviroments -Decreased locomotion in familar enviroments | ND in grooming and rearing behaviors | -ND in anxiety for OFT, EPM and Light-Dark box | | Spontaneous seizures -ND in susceptibility to PTZ-induced seziures | ND for novel object recognition |
| | | | 4 days - 3 weeks | -ND in number and duration of USVs -Slight increase direct social interaction | | Decreased locomotion OFT | ND in grooming | ND OFT | | | ND in mother attachment behaviors |
| **Hypomorphic Scn2a Mouse Model Behavior** ||||||||||||
| Wang et al., 2021 | Scn2a Δ1898/+ | | >= 30 to 44 days | -Males displayed increased social interacton with novel mouse in 3-chamber - After repeat exposure to the same mouse, males , still displayed increased social interaction | | ND in general locomotion | ND in time spent grooming | -Increased activity in OFT - Reduced anxiety-like behaviors on EPM | | | |
| Eaton et al., 2021 | Scn2a gtKO/gtKO (AKA Scn2a Trap) | | 3 to 5 months | | -Decreased novel object -Increased fixed-pattern of exploration in novel object -Decreased number of entries to, duration in, and prefrence for the novel arm of Y-maze | -ND activity, tail posture, limb grasping, extension, righting reflex -Decreased grip ability in SHIPRA -Females traveled longer -Decrease in distance traveld in OFT perfromed with increased birightness | -Decreased overall nest building -Decreased quality of nests built -Increased grooming -Slightly increased induced marble burying | -Increased anxiety-like behavior in low and bright light OFT -ND freezing behavior in low OFT -Increased freezing in OFT perfromed with high intensity lumnesience | -ND overall odor discrimination -Increased in time for discrimination | | -Increased water and food intake and number of fecal boli in males -Lower body position rating, less trunk curl and postural reflex when held by SHIPRA -Decresed tolerance for heat and cold temperatures on themerical/cold assay -ND for dynamic hot plate Increased latency -Decreased startle response in PPI |
| Ma et al., 2022 | | | > 6 weeks | | | Impairment of wheel-running activity in light-dark conditions | | | | | Abnormal sleep patterns |
| Wu et al., 2024 | | | 3 to 4 months | | Impaired learning and memory on Borris water maze and Y-maze | | | | | | |
| **Conditional and other Mouse Models of Scn2a Insufficiency** ||||||||||||
| Wang et al., 2024 | Scn2a +/fl | Alpha6-Cre | 60 to 90 days | | | | | | | | Altered VOR |
| Tatsukawa et al., 2019 | Scn2a fl/+ | Emx1-Cre | 8 to 9 weeks (Male Only) | | | Increased rearing in OFT | | -Increased time in center of OFT -Trend of entering open arms of EPM more | | | |
| | | Vgat-Cre | | | | ND in locomotion or rearing | | -ND OFT Reduced anxiety-like behaviors on EPM | | | |
| Miyamoto et al., 2019 | | Emx1-Cre | >= 8 weeks | | | | | | | -SWDs present during behavioral quiescence -Inactivation of the CPu or thalamus by muscimol injection suppresses SWDs | |
| | | Vgat-Cre | | | | | | | | No spontaneous SWDs | |
| | Scn2a fl/fl | Trpc-Cre | | | | | | | | Spontaneous SWDs | |
| | | Ntsr-Cre | | | | | | | | No spontaneous SWDs | |
| Ma et al., 2022 | Scn2afl/fl | SCN-Cre | 3-5 months | | | | | | | | Sleep Disturbances |
| Suzuki et al., 2023 | Scn2a fl/fl | mPFC-Cre | 18 to 19 weeks | Increased sociablity in three chamber social approach | | -Less distance traveld in chamber social assay -Less distance travelled and fewer rearing events in OFT | | Increased anxiety in OFT | | | Reduced PPI |
| | | VTA-Cre | 11- 13 weeks | ND sociabillity in 3 chamber social approach | | Reduced rearing in OFT | | ND OFT | | | Increased percentage of PPI of startle response to 120 dB sound with 74 or 78 dB prepulse |
| Nunes and Kuner, 2018 | Scn2a :OB GC -shRNA | | 49 to 63 days | | | | | | Reduced speed in odor discrimination | | |

| Paper | Model | Iteration of Conditional Model | Mice Age | Social | Learning/Memory | Motor | Repetitive Behaviors | Anxiety/Hyperactivity | Olfaction | Seizures | Other |
|---|---|---|---|---|---|---|---|---|---|---|---|
| Li et al., 2023 | *Scn2a ASO* | | 2-3 months | -Impaied social approcah say <br> - Increased social avoidiance in 3-chamber assay | | Increased locomotion in OFT | Impaired nest building behavior | -Decreased aniexty and ability to assess risk in in EPM - Increased anxiety-like behaviros in OFT | | -No spontaneous seziures or detecable ictal features - Augmented seziure response to PTZ treatment | |